\documentclass[11pt,twoside]{article}
\usepackage{FUSE2004}
\usepackage{natbib}

\usepackage{epsf}
\usepackage{psfig}
\usepackage{lscape}

\begin{document}
\title{Intrinsic Absorption in NGC 7469}
\author{Jennifer E. Scott  and the FUSE AGN Working Group}
\affil{Space Telescope Science Inst., 3700 San Martin Dr.,
Baltimore, MD  21218}

\begin{abstract}
We present results from a coordinated {\it FUSE},
{\it HST}/STIS and {\it Chandra} campaign
to study intrinsic UV and X-ray absorption in the
outflow of the Seyfert 1 galaxy NGC~7469.  
Previous non-simultaneous observations of
this outflow found two distinct UV absorption components, one of which
likely corresponds to the X-ray absorber.  The {\it FUSE} data reveal that
the \ion{O}{6} absorption in this component has strengthened over time,
as the continuum flux decreased.
We use  measured
\ion{H}{1}, \ion{N}{5}, \ion{C}{4}, and \ion{O}{6} 
column densities to model self-consistently the
photoionization state of the absorbers. 
We confirm the physical picture of the outflow in which the low velocity component
is a highly ionized, high density absorber located near the broad emission line region, 
while the high velocity component 
is of lower density and resides farther from the central engine.
\end{abstract}

\section{Introduction}
Absorption edges are visible in the X-ray spectra of
about one half of all low redshift AGN (Reynolds 1997;
George et al.\ 1998; Crenshaw, Kraemer \& George 2003),
virtually all of which also show
high-ionization absorption lines in their UV spectra (Crenshaw
et al.\ 1999),
suggesting a connection between
the two phenomena.
The absorption is thought to be intrinsic to the AGN because
the complexes are generally
blueshifted with respect to the AGNs,
and because many absorbers show variability 
and/or non-unity covering fractions. (See references in Kriss et al.\
in this proceeding.)

Most high resolution observations of the
intrinsic absorption in AGNs,
with the exception of recently published observations of
Mrk~279 (Scott et al.\ 2004), have
been performed at different times in the UV and X-rays.
The high degree of variability in AGN continua
complicates self-consistent photoionization modeling
from these non-simultaneous data, preventing firm
conclusions about the nature of the relationship between UV and X-ray
absorbers.

The intrinsic absorption in NGC~7469 has been studied previously
using UV and X-ray observations separated by one year (Blustin et al.\ 2003;
Kriss et al.\ 2003).
These authors found two primary UV components, with outflow
velocities of -569 and -1898 km s$^{-1}$, Components 1 and 2, respectively.  
The {\it Chandra (CXO)}, {\it FUSE}, and {\it HST}/STIS campaign presented here
is the first set of simultaneous, high-resolution
UV and X-ray spectral observations of NGC~7469.
Here, we describe the UV and X-ray observations and 
the properties of the AGN outflow, and we
discuss the geometry of the absorption
using the results from photoionization models.

\section{Data and Analysis}
We obtained a 150 ksec {\it CXO}/HETG spectrum of NGC~7469 on
2002 December 12-13, covering 0.5-10 keV.
We fit this with an absorbed blackbody + power law continuum,
with a  photon index of 1.82 and Galactic $N_{H}=4.87 \times 10^{20}$ cm$^{-2}$.
On 2002 December 13, we observed NGC~7469 with the
$30\arcsec \times 30\arcsec$ low-resolution aperture of
{\it FUSE}, obtaining a total exposure of 7 ksec.
Simultaneously, we obtained 13 ksec of STIS data
using the E140M grating and the $0.2\arcsec \times 0.2\arcsec$ aperture
covering 1150-1730~\AA.
We fit the {\it FUSE} and STIS data with a single power law continuum:
$F_{\lambda}=(8.18 \pm 0.06) \times 10^{-14}$ ($\lambda$/1000~\AA)$^{-1.082 \pm 0.021}$
ergs s$^{-1}$ cm$^{-2}$ \AA$^{-1}$ and with Gaussian emission line profiles
from \ion{O}{6}, Ly~$\alpha$, \ion{N}{5}, \ion{O}{1} + \ion{S}{3},
\ion{C}{4}, \ion{He}{2}, \ion{Si}{2}, and \ion{Fe}{2} in the vicinity of the
intrinsic absorption lines of interest.

We also observed NGC~7469 with STIS on 2004 June 21-22
for 23 ksec using the same setup described above.
For this spectrum, we fit
$F_{\lambda}=(3.25 \pm 0.01) \times 10^{-14}$ ($\lambda$/1000~\AA)$^{-1.086 \pm 0.007}$
ergs s$^{-1}$ cm$^{-2}$ \AA$^{-1}$.

\begin{figure}
\plottwo{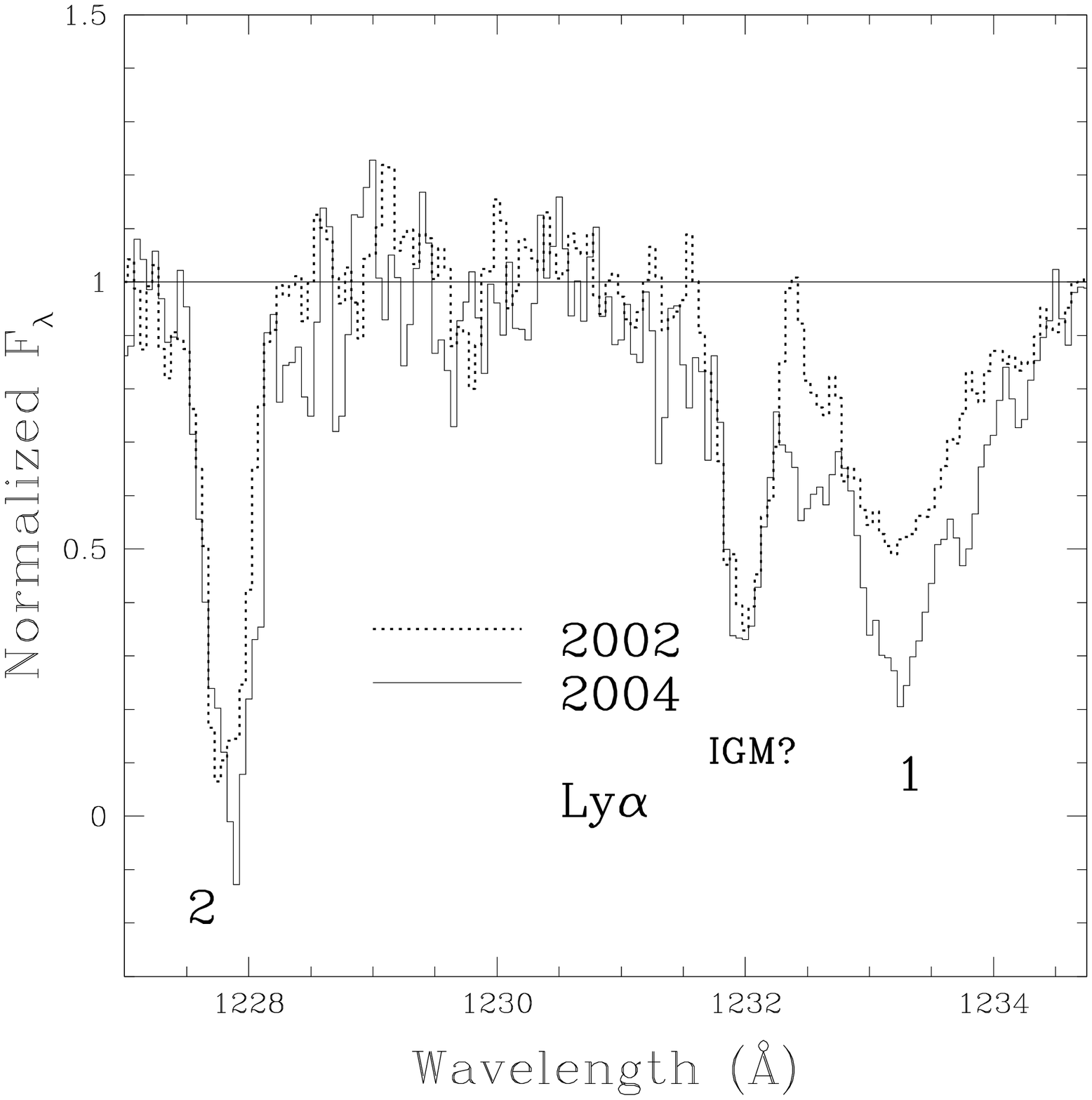}{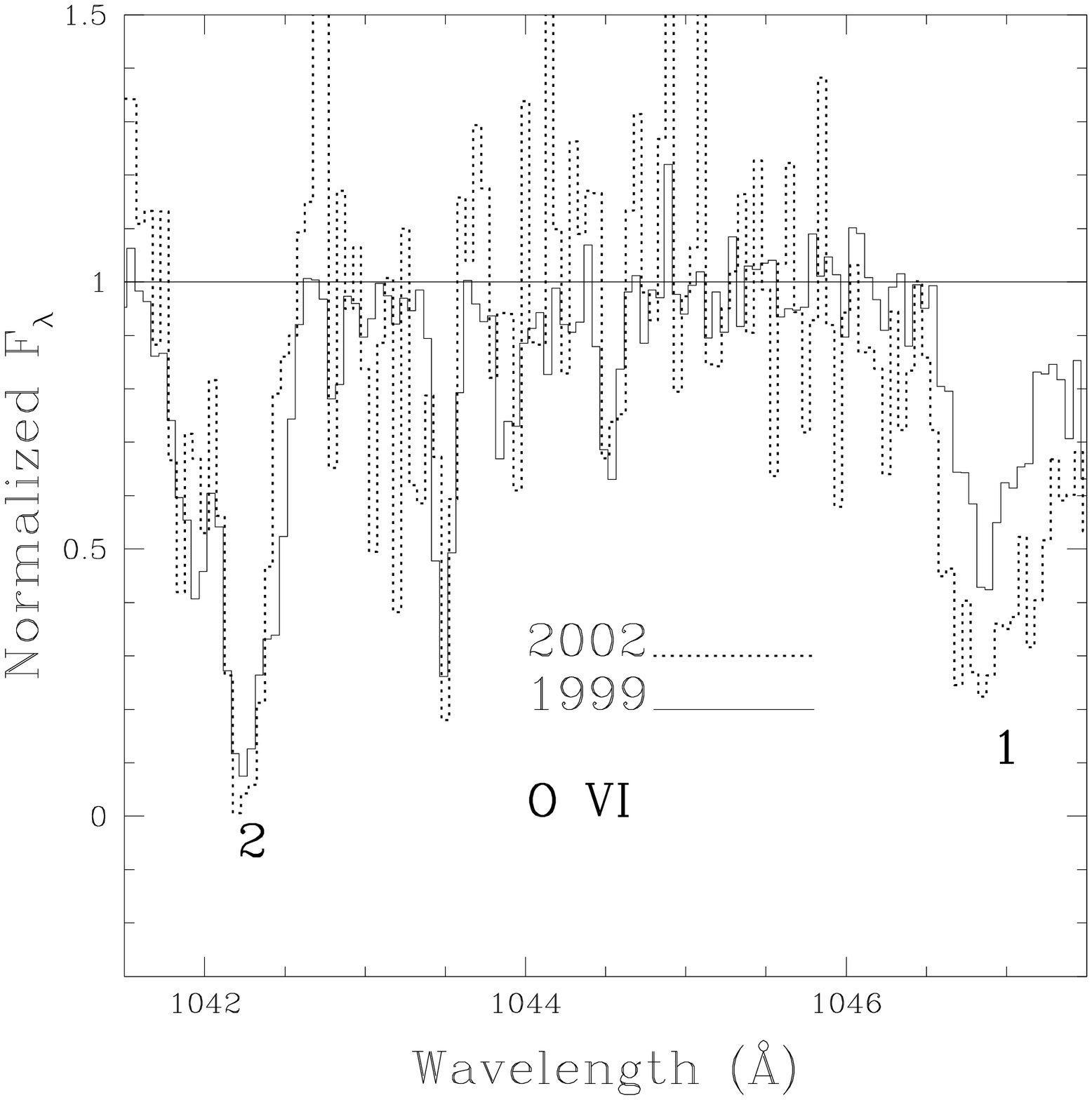}
\caption{Normalized profiles of Ly$\alpha$ and \ion{O}{6}
absorption
over two epochs of STIS and {\it FUSE} observations with
Components 1 and 2 marked.}
\end{figure}

We measured covering fractions and column densities of the intrinsic absorbers
using the IRAF task {\it specfit} (Kriss 1994), and
we compared photoionization models (Krolik \& Kriss 1995, 2001)
to the measured \ion{H}{1}, \ion{N}{5}, \ion{C}{4}, and \ion{O}{6}
column densities.
We incorporated the observed
X-ray and UV slopes
and normalizations at the time of the 2002 observations
into the input spectral energy distribution for NGC~7469 (Kriss et al.\ 2000, 2003).

\section{Results and Conclusions}
We summarize our results as follows:

\begin{enumerate}
\item{Component 1 shows saturated or nearly saturated \ion{O}{6} absorption.
Neutral hydrogen is clearly present, as Ly$\alpha$ is a strong feature, but 
Ly$\beta$ is not detected.  The {\it FUSE} spectrum near  Ly$\beta$ is
heavily contaminated by molecular hydrogen absorption.  The \ion{N}{5} and
\ion{C}{4} features associated with this velocity component are weak.}  

\item{Component 2 shows prominent \ion{O}{6} and Ly$\alpha$ absorption.
As for Component 1, no Ly$\beta$ is found, but the \ion{N}{5} and
\ion{C}{4} features are stronger than those of Component 1.} 

\item{We do not detect \ion{O}{7} or \ion{O}{8} absorption in the
{\it CXO} spectrum, but we do
detect \ion{O}{8} emission and absorption from H-like and He-like
Si, Ne, and Mg
at velocities consistent with Component 1.
Component 2 has no associated X-ray absorber.}

\item{Consistent with previous results (Kriss et al.\ 2003), the 
covering fraction for Component 1 is $\sim$0.5, while that of Component 2
is consistent with one.}
\item{We see absorption variability in Component 1 over the three 
observation epochs (Fig.\ 1): 
the \ion{O}{6} absorption strengthened between the 1999 {\it FUSE} observations
(Kriss et al.\ 2003) 
and those presented here,
as the continuum flux decreased by a factor of $\sim$1.5; and
the Ly$\alpha$ absorption increased between 2002 and 2004 as the
continuum flux decreased by a factor of $\sim$2.5.}
\item{The photoionization models combined with the
\ion{H}{1}, \ion{N}{5}, \ion{C}{4}, and \ion{O}{6}
column density measurements
show that (log(N),U)$\sim$(20.0,1.0) 
for Component 1 and (log N,U)$\sim$(18.5,0.08)
for Component 2. The \ion{C}{4} and  \ion{N}{5} column densities are
particularly useful for breaking the degeneracy in models based on
\ion{O}{6} alone.}
\end{enumerate}

Our conclusions are
consistent with the previous physical picture of the outflow 
(Blustin et al.\ 2003; Kriss et al.\ 2003).
Component 1 is a highly-ionized, high density absorber that is
coincident with or interior to the broad emission line region and 
related to the X-ray absorption and emission component. 
Compared with Component 1, 
Component 2 is a lower density and ionization parameter system
located at a larger physical distance from the central engine
in the absorbing outflow.

% For using the "thebibliography" environment use these.
% See the "Authors Instructions" for details.

\end{document}